\documentclass[%
 aip,
 amsmath,amssymb,
 reprint,%
]{revtex4-2}

\usepackage{graphicx}% Include figure files
\usepackage{dcolumn}% Align table columns on decimal point
\usepackage{bm}% bold math
\usepackage[utf8]{inputenc}
\usepackage[T1]{fontenc}
\usepackage{mathptmx}
\usepackage{etoolbox}

\makeatletter
\def\@email#1#2{%
 \endgroup
 \patchcmd{\titleblock@produce}
  {\frontmatter@RRAPformat}
  {\frontmatter@RRAPformat{\produce@RRAP{*#1\href{mailto:#2}{#2}}}\frontmatter@RRAPformat}
  {}{}
}%
\makeatother
\begin{document}

%\preprint{AIP/123-QED}

\title[]{Quantum Treatment of the Current through Plasma-Metal Junction: Fundamentals}
% Force line breaks with \\
\author{Muthukumar Balasundaram}
\homepage{https://www.pondiuni.edu.in/faculy_profiles/dr-b-muthukumar/}
 \email{muthukbs@pondiuni.ac.in}
\author{Suraj Kumar Sinha}%
\homepage{https://www.pondiuni.edu.in/faculy_profiles/dr-suraj-kumar-sinha/}
% \email{sinhasuraj.phy@pondiuni.edu.in}
\affiliation{Department of Physics, Pondicherry University, Kalapet, Pondicherry-605014, India.%\\This line break forced with \textbackslash\textbackslash
}%

\date{\today}% It is always \today, today,

\begin{abstract}
We study the quantum nature of current through plasma-probe junction from the viewpoint of the metal probe. The intrinsic material properties of the metal and their influence on the nature of the observed current are theoretically worked out. The novel idea is that the plasma-sheath at the plasma-probe junction is treated as a potential barrier, and in analogy with the current conduction through a metal-metal junction, the current through the plasma-sheath is treated as a quantum barrier penetration problem.  Essentially, we obtain an expression for the electron-current as a function of the bias voltage in its full range, thereby unlocking the intricate dependency of the current on the material properties of the probe.
\end{abstract}

\maketitle

\section{Introduction \label{sec-intro}}
When an electrically isolated metal is brought into contact with a plasma, the electrons in the metal and in the plasma attain a thermal equilibrium. This phenomenon can be explained based on a key idea introduced in \cite{2020PhPl...27b3512A}. The idea in \cite{2020PhPl...27b3512A} attributes a Fermi energy level to the plasma electrons, but in this article we adopt a different interpretation to that energy level since the plasma electrons are not considered to obey the Fermi-Dirac statistics. At the plasma-metal (PM) junction, where quasi-neutrality of plasma-ions and electrons is lost due to much more mobile electrons, the surface of the metal tends to get negatively charged. 

Any accumulated  charge on a conductor  creates an electrostatic potential inside the conductor, and this potential must be constant throughout the metal conductor in order for electrostatic equilibrium to be possible \cite{2009mtem.book.....J}. For a negatively charged metal that is placed in an air medium, the constant value of such a potential is equal to its value at the surface \cite{2009mtem.book.....J}. The effect of this surface charge is to increase the amount of work done to extract an electron from the metal surface \cite{1958.book.Ehrenberg}. 
\section{Floating Metal\label{sec-float}}
In the following we consider the metal probe to be a one-dimensional rod coated by a thick dielectric material on its cylindrical sides, and the only direction at which it makes direct contact with the plasma is at a tip where it is not covered by the dielectric material. The conduction of electrons is assumed to take place either way only through this tip. The plasma is assumed to be non-magnetized, cold and weakly ionized obeying the Maxwellian velocity distribution. Since plasma is an electrically conducting medium, the electrons in the metal and in the plasma can tunnel through the sheath barrier at the tip and attain an electrical and thermal equilibrium state. Therefore, following \cite{2020PhPl...27b3512A}, we propose the following ideal energy-band diagram for the plasma-metal junction at equilibrium in the case of the floating metal with all the electrons at the absolute zero temperature:
\begin{figure}[ht]
\begin{center}
\includegraphics[scale=0.23]{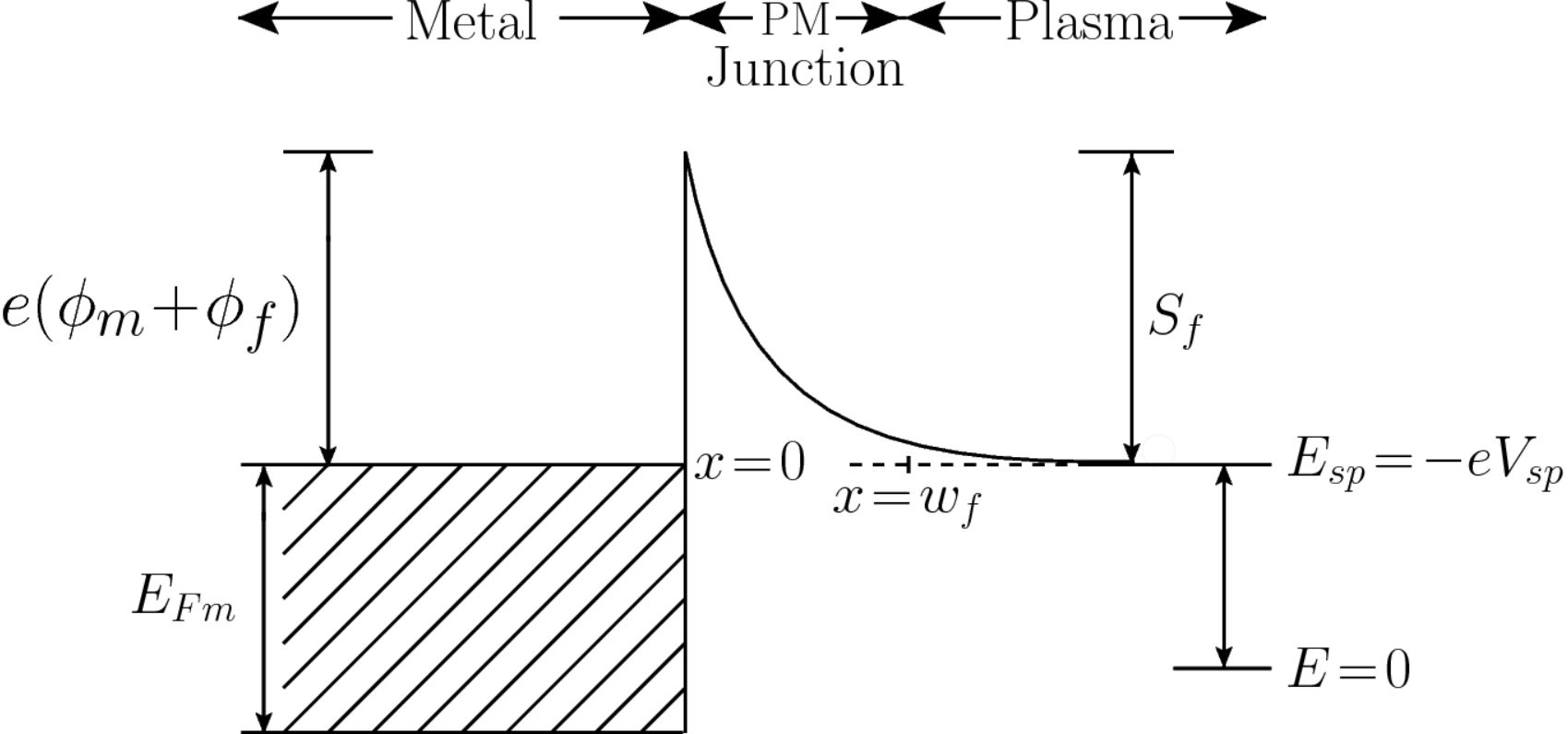}%
\caption{Equilibrium configuration of floating metal\label{pic-float}}
\end{center}
\end{figure}
The surface charge density is actually more at the tip  and so the conduction band of the metal is bent downward with a downward-cusp inside the metal at  $x=0$ (see for example \cite{1958.book.Ehrenberg}). Since negative charges tend to get settled down even along the rod above the dielectric coating, the surface along the rod also tends to get charged and so we are approximating the conduction band as a straight band at $x=0$ for the ease of later calculations. Therefore, the work function $e\phi_m$ of the floating metal is taken to pick-up the same correction $e\phi_f$ all along the rod. The effective work function $e(\phi_m+\phi_f)$ should be seen as the work done in extracting a metal electron from the surface of the metal. 

The plasma (space) potential $V_{sp}$ and the energy line $E=0$ are chosen in such a way that an electron at rest in the plasma has the potential energy $E_{sp}=-eV_{sp}$ which is the minimum energy a plasma electron can have. To occupy any energy level lower than this potential energy, a plasma electron will have to have an imaginary momentum which is unphysical. At equilibrium, the Fermi energy level corresponding to the Fermi energy $E_{Fm}$ of the metal should be in line with the space potential energy line $E_{sp}$. This is because if the Fermi level of the metal is lower than the space potential energy line, then the electrons on the plasma side would tunnel through the barrier into the metal to occupy a lower energy state, and if the space potential energy line is lower than the Fermi level of the metal, then the metal electrons would tunnel through the barrier to the plasma side to occupy a lower energy state, and this process will continue until the equilibrium state is achieved. When the Fermi level of the metal and $E_{sp}$ are aligned, on the one hand, it is the Pauli's exclusion principle that would prevent a plasma electron  to the metal side, and on the other hand, the minimum energy condition on the plasma side would prevent a metal electron to come to the plasma side. 

The alignment of the energy levels shown in Fig.\ref{pic-float} makes it easy to understand the nature of $\phi_{f}$ inside the floating metal. An electron at the top of the conduction band in the metal has to surmount both the work function $e\phi_{m}$ and the correction  $e\phi_{f}$ to jump over the sheath barrier of height $S_f$ to the plasma side. Therefore, the sheath potential $S_f$ at the tip of the floating metal is given by 
%
%\begin{align}
$S_f(x=0)=e(\phi_m+\phi_f).$ 
%\end{align}
%
The width of the Debye sheath at the tip of the metal in this floating case is denoted by $w_f$. With this background, we will set up the condition for equilibrium. If $J_{mp}^{e-tr}$ denotes the transmitted current of electrons from metal to plasma, $J_{pm}^{e-tr}$ the transmitted current from plasma to metal, $J_{mp}^{e-re}$ the recombination current of the electrons that neutralize any impinging ion on the surface of the metal tip and $J_{mp}^{e-se}$ the current of the secondary emission of electrons, then the equilibrium condition is that the number of metal electrons should remain the same. This is achieved if
%
%\begin{align}
$J_{pm}^{e-tr}=J_{mp}^{e-tr}+J_{mp}^{e-re}+J_{mp}^{e-se}$. 
%\end{align}
%
If we assume that all the ions are singly ionized and that they are neutralized only upon impinging on the tip of the metal and not when they pass through the sheath, then $J_{mp}^{e-re}=J_{pm}^{i}$, where $J_{pm}^{i}$ is the ion-current from plasma to metal. Since the secondary emission of electrons is the result of neutralization of ions, we have $J_{mp}^{e-se}=\gamma_{se} J_{pm}^{i}$ \cite{2005.book.Liberman}, where $\gamma_{se}$ is the secondary emission coefficient. Then the condition for equilibrium becomes
\begin{align}
\displaystyle J_{pm}^{e-tr}\!=\!J_{mp}^{e-tr}+(1+\gamma_{se})J_{pm}^{i}. \label{eq-eqbm-cond1}
\end{align}
Since the probe is assumed to be one-dimensional,  the current $J_{mp}^{e-tr}$ can be written as
\begin{align} \displaystyle
 J_{mp}^{e-tr}\!=\!-2\,e\!\!\int\! dE_m \, f_m(E_m)\, g_m(E_m) \left(\dfrac{p_m}{m_e} \right) D_{mp}^f, \label{eq-mp-ecurrent1}
\end{align}
where $-e$ is the electron charge, $m_e$ is the mass of the electron, $p_m$ is the momentum of the metal electron, $g_m(E_m)$ is the number of electronic states per unit length per unit energy interval in one dimension without spin, $D_{mp}^f$, which is a function of $E_m$, denotes the probability for a metal electron with energy $E_m$ to penetrate through the sheath barrier from metal to plasma, $f_{m}(E_m)$ is the probability that a metal  electron state of energy $E_m$ is occupied and it is given by   
\begin{align}
 f_m(E_m)=\left\{ 1+\exp[{(E_m-E_{Fm})/KT_{me}}]\right\}^{-1} \label{eq-fd-distri},  
\end{align}
where $T_{me}$ is the temperature of the metal electrons. The factor $2$ in Eq.\eqref{eq-mp-ecurrent1} accounts for the spin. The density of states per unit length in the one-dimensional case is given by $g_m(E_m)=m/(2\pi\hbar^2k_m)$ \cite{1969Kane}, where $k_m=p_m/\hbar$. Therefore, Eq.(\ref{eq-mp-ecurrent1}) reduces to     
\begin{align}
 J_{mp}^{e-tr}=-\frac{e}{\pi\hbar}\int dE_m \, f_m(E_m) \,D_{mp}^f(E_m). \label{eq-mp-ecurrent2}
\end{align}
Since the metal part is treated as  one dimensional, we will treat the plasma part also as one dimensional.  The current $J_{mp}^{e-tr}$ can then be written as
\begin{align}
 J_{pm}^{e-tr}=-e\!\!\int \!dv_p \, v_p\,f_p(v_p)\, [1-f_m(E_p)] \,D_{pm}^f(E_p), \label{eq-pm-ecurrent1}
\end{align}
where $f_p(v_p)$ is the number of electrons per unit length per unit velocity, $(1-f_m(E_p))$ is the probability that the metal electron state with energy $E_p$ is unoccupied so that a plasma electron with energy $E_p$ can penetrate through the sheath-barrier and occupy that state. $D_{pm}(E_p)$ is the transmission probability for such penetration to take place from plasma to metal. In terms of energy integration, Eq.(\ref{eq-pm-ecurrent1}) can be written as
\begin{align}
 J_{pm}^{e-tr}=-\frac{e}{m}\int dE_p \,f_p(E_p)\, [1-f_m(E_p)] \,D_{pm}^f(E_p). \label{eq-pm-ecurrent2}
\end{align}
We write the ion-current in the following way:
\begin{align}
 J_{pm}^{i}=e\int dv_p^i \, v_p^i\,f_p(v_p^i) \,D^i_f(E_p^i), \label{eq-ion-current}
\end{align}
where $dv_p^i\,f_p(v_p^i)$ is the number of ions per unit length having velocities from $v_p^i$ to $v_p^i+dv_p^i$, and $D^i_f(E_p^i)$ is the probability for an ion with energy $E_p^i$ to penetrate through the sheath barrier to hit the tip of the floating metal. Compared to the value of the electron current, the ion current is small and for the case of the ion-temperature close to absolute zero, we assume that the ion current is negligibly small i.e., $J_{pm}^i \approx0$. So the equilibrium condition Eq.(\ref{eq-eqbm-cond1}) becomes   
\begin{align}
 J_{pm}^{e-tr}=J_{mp}^{e-tr}. \label{eq-eqbn-float}
\end{align}
\section{Biased Metal at Floating Potential\label{secbiasfloat}}
\begin{figure}[ht]
\begin{center}
\includegraphics[scale=0.23]{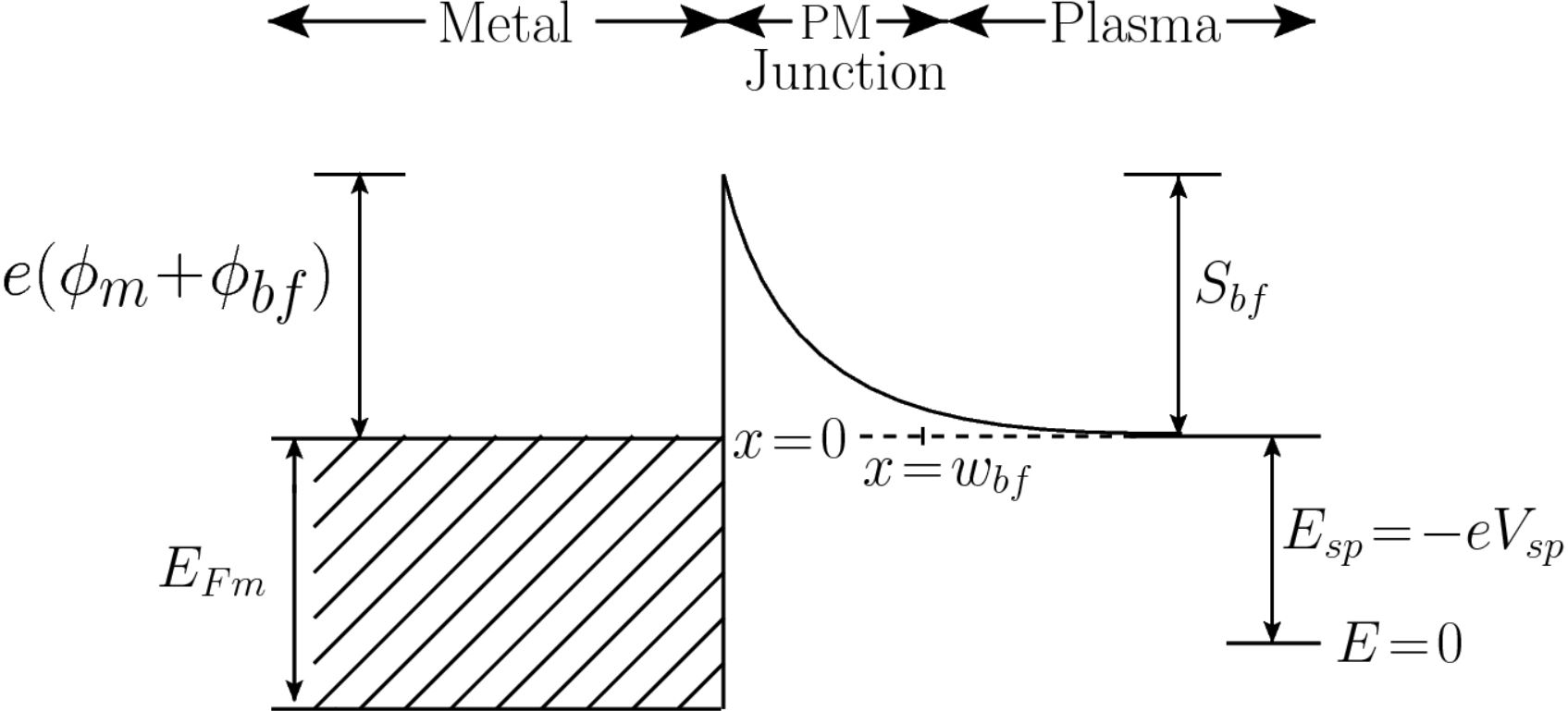}%
\caption{Biased metal at floating potential} \label{pic-biasfloat}
\end{center}
\end{figure}

If a potential $V_m$ is applied to the metal, it is given the bias $V_b=V_m-V_{sp}$ with respect to the space-potential. At the floating potential $V_m=V_f$,  there is no net current flowing in either direction. We call this state the equilibrium state at floating potential. For this state, the energy-band diagram is depicted in Fig.\ref{pic-biasfloat}. The suffix in the quantities $\phi_{bf}$, $S_{bf}$ and $w_{bf}$ denotes the biased case at floating potential $V_m=V_f$. These values are respectively different from $\phi_f$, $S_f$ and $w_f$, but the interpretation of these quantities are the same as in the floating metal case. 
The currents Eq.(\ref{eq-mp-ecurrent2}) and Eq.(\ref{eq-pm-ecurrent2}) are written in this case as
\begin{align}
 J_{mp}^{e-tr}=-\frac{e}{\pi\hbar}\int dE_m \, f_m(E_m) \,D_{mp}^{bf}(E_m), \label{eq-mp-ecurrent-bf}
\end{align}
\begin{align}
 J_{pm}^{e-tr}=-\frac{e}{m}\!\int dE_p \,f_p(E_p)\, [1-f_m(E_p)] \,D_{pm}^{bf}(E_p), \label{eq-pm-ecurrent-bf}
\end{align}
and in this case also the above currents obey the equilibrium condition under the same assumptions as in the floating metal case:
\begin{align}
 J_{pm}^{e-tr}=J_{mp}^{e-tr}. \label{eq-eqbn-floatpot}
\end{align}
At this point, we apply the flat-top approximation in which the sheath potential is treated as a square-barrier. 
\begin{figure}[ht]
\begin{center}
\includegraphics[scale=0.23]{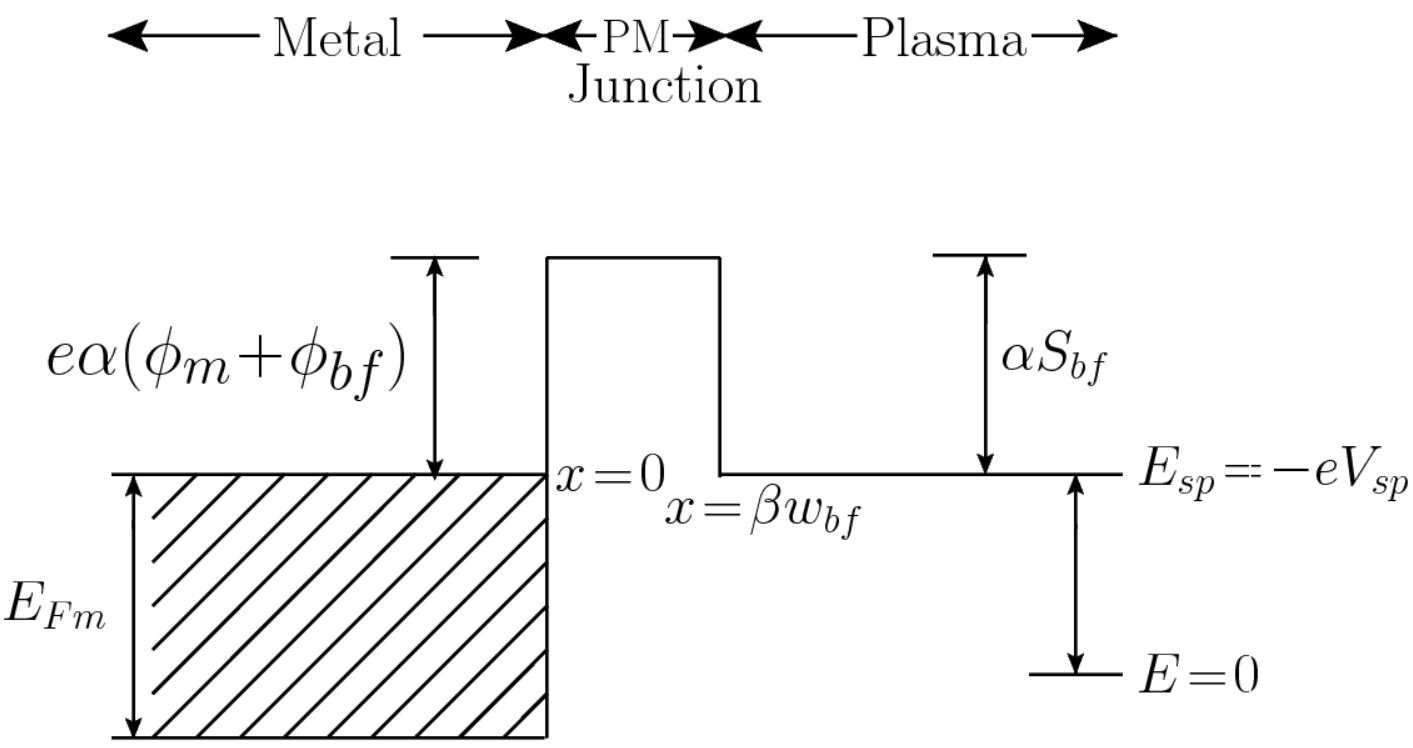}%
\caption{Biased metal at floating potential - the square barrier model\label{pic-flattopbiasfloat}} 
\end{center}
\end{figure}
Such a square-barrier approximation has already been applied for the case of metal-barrier-metal tunneling in the three dimensional case \cite{1969Duke}. The key point in \cite{1969Duke} is that the shape of the effective square-barrier depends
explicitly on the bias leading to an elegant description of current-voltage characteristics. In our case, we take the height and the width of the square-barrier respectively to be $\alpha e(\phi_m+\phi_{bf})$ and $\beta w_{bf}$, where $\alpha$ and $\beta$ are some scale factors to be fixed later. 
%which can be fixed later to match the current-voltage characteristics with the experimental results. 
Assuming that there are no radiative losses across the barrier, our PM-junction case is just another barrier-penetration problem as in \cite{1969Kane,1969Duke,1930PhRv...36.1604F}.  
%However there are two main differences in our case: 1) the ways in which the height and width of the barrier depend on the bias voltage. But before going into those details, 
It is worth pointing out that for such barrier-penetration problems, the transmission probabilities are independent of the direction of transmission across the barrier \cite{1930PhRv...36.1604F,1969Kane,1969Duke}, i.e., $D_{pm}^{bf}(E_p)=D_{mp}^{bf}(E_p)$. Therefore, the condition Eq.(\ref{eq-eqbn-floatpot}) together with the law of conservation of the total energy, $E_p=E_m$, imply that
\begin{align}
 f_p(E_p)=\left(\frac{m}{\pi\hbar}\right)\left(\frac{f_m(E_p)}{1-f_m(E_p)}\right). \label{eq-eqbm-distrifn}
\end{align}
The key differences between the metal-metal junction and PM-junction are the following:  1) In the case of a metal in contact with another metal, at the boundary there is a small gap, corresponding to the vacuum level, which acts like a barrier and has a finite width. Also, the barrier has a flat-top in the unbiased case. In the case of plasma-metal junction, since the plasma is not a solid material, there is no such gap. When an electron comes out of the metal overcoming the work function, it immediately encounters the sheath potential and not any vacuum level. 2) Both the height and the width of the sheath barrier depend on the bias voltage. 
\section{Biased Metal\label{sec-bias}}
In the biased metal probe case, we note the following points for the probe potential $V_m$: 

\begin{enumerate}
 \item For $V_m<V_f$, the ions are the attracted particles leading to the formation of ion-sheath around the probe and the electron current is saturated \cite{raizer2011gas}. 
 \item At the floating potential $V_m=V_f$, we have $\phi_b=\phi_{bf}$.
 \item At $V_m=V_{sp}$, there is no sheath \cite{conde2011introduction}.
 \item Beyond $V_{sp}$, the ions are repelled by the probe, and a layer of negative sheath develops around the probe to screen the excess potential $V_m-V_{sp}$ \cite{raizer2011gas}.
\end{enumerate}
\begin{figure}[ht]
\begin{center}
\includegraphics[scale=0.23]{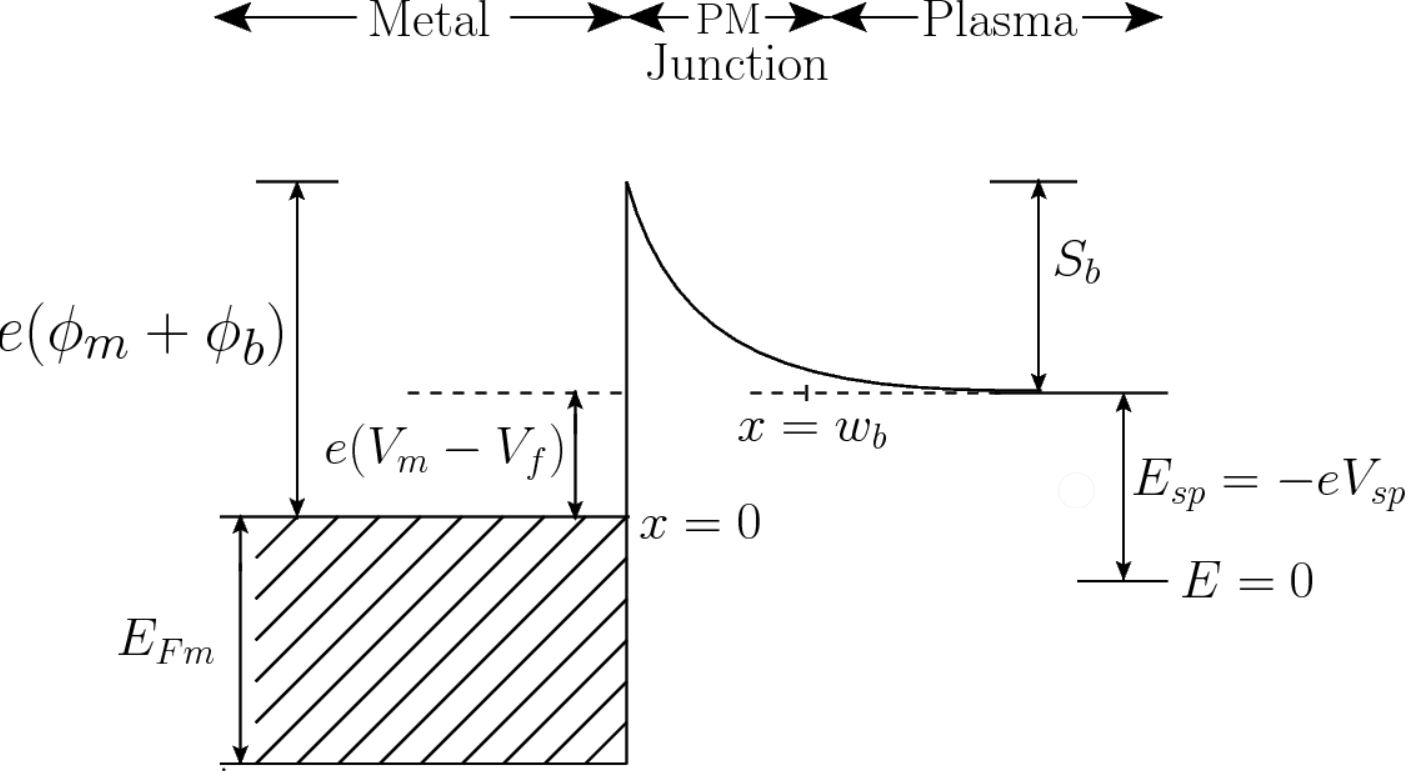}%
\caption{A Positive bias $(V_m-V_f)$ applied on the metal relative to the floating potential $V_f$ pushes its Fermi level vertically down.\label{pic-bias}}
\end{center}
\end{figure}

\begin{figure}[ht]
\begin{center}
\includegraphics[scale=0.23]{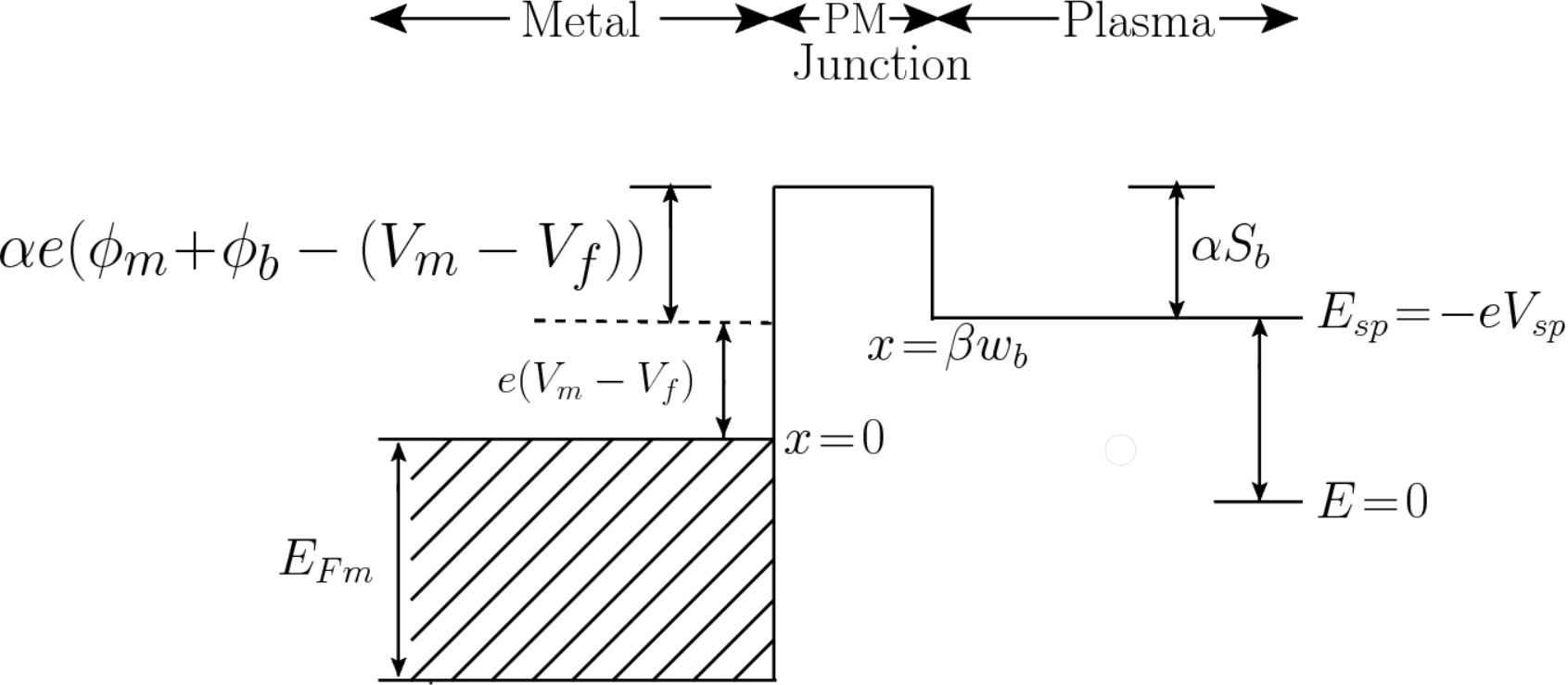}%
\caption{A Positive bias $(V_m-V_f)$ applied on the metal relative to the floating potential - the square barrier  model.\label{pic-biasflattop}}
\end{center}
\end{figure}
In Fig.\ref{pic-bias} we take $\phi_b$ in such a way that $e(\phi_m+\phi_b)$ is the work needed to extract a metal electron from the top of the conduction band to just outside the metal where the sheath potential $S_b$ starts acting on the electron. Therefore, for the potential $V=V_m-V_f$ relative to the floating potential, we have
%
%\begin{align}
$S_b=e(\phi_m+\phi_b-V)$. 
%\end{align}
%
The height of the effective square barrier above the $E_{sp}$ line is taken to be $\alpha S_b$ and the width of this barrier is taken to be $\beta w_b$. Following the convention of $+$ve sign for the electron current from plasma to metal, the net current is written as
\begin{align}
J=J_{mp}^{e-tr}-J_{pm}^{e-tr} \label{netCurrent}, 
\end{align}
where
\begin{align}
 J_{mp}^{e-tr}=-\frac{e}{\pi\hbar}\int dE_m\, f_m(E_m+eV)\,D^b(E_m), 
\end{align}
and
\begin{align}
 J_{pm}^{e-tr}=-\frac{e}{m}\int dE_p \,& [1-f_m(E_p+eV)] \times\nonumber \\ & \times f_p(E_p)\,D^b(E_p). \label{eq-bias-pm-current}
\end{align}
In the above expressions, $D^b(E)$ denotes the transmission probability in the case in which a bias $V=V_m-V_f$ is applied to the metal relative to the floating potential $V_f$. 
Substituting the equilibrium condition Eq.(\ref{eq-eqbm-distrifn}) in Eq.(\ref{eq-bias-pm-current}), we arrive at the following expression for the net current
\begin{align}
 J\!=\!\frac{e}{\pi\hbar}\!\left(\!\exp{\!\left[\frac{eV}{KT_{me}}\right]\!}\!-\!\!1\!\right)\!\!\int\limits_{E_{sp}}^{\infty}\! dE\,f_m(E+eV)\, D^b(E), \label{eq-biased-J}
\end{align}
where $E=E_p=E_m$ due to conservation of energy. 
\section{Energy and Transmission Probability of Electrons}
For plane waves in the plasma region, we have the following energy relationship:
\begin{align}
 E_p-E_{sp}&= \frac{\hbar^2 k_p^2}{2m_e} \label{eq-plasma-e-energy}
\end{align}
since the plasma particles are assumed to be free.  In the metal region, plane waves have the energy relation 
\begin{align}
% (\mathrm{plasma~particles~are~assumed~to~be~free}),\\
 E_m+(E_{Fm}-E_{sp})&=\frac{\hbar^2 k_m^2}{2m_e} \label{eq-metal-e-energy} \quad %(\mathrm{parabolic~band~model~for~a~metal~ electron}).
\end{align}
since the parabolic band model is used for a metal electron, where $(E_{Fm}-E_{sp})$ is the additional energy the electron acquires due to the energy measurement from the bottom of the conduction band of the metal. 

From Fig.\ref{pic-biasflattop}, it is clear that a plasma electron sees the barrier of height $\alpha S_b=\alpha e(\phi_m+\phi_b-(V_m-V_f))$. Therefore, the energy of an electron in the sheath-barrier can be written as

\begin{align}
 E_b=\frac{\hbar^2 k_b^2}{2m_e} +\left[E_{sp}+\alpha e(\phi_m+\phi_b-(V_m-V_f))\right] \label{eq-trans-energy-sq-barrier}.
\end{align}
The conservation of energy implies that
\begin{align}
 E_p=E_m=E_b, \label{eq-energy-conserve}
\end{align}
which will be denoted as $E$ in the subsequent analysis. 

For the effective square-barrier approximation shown in Fig.\ref{pic-biasflattop}, the transmission probability $D^b(E)$ is given by \cite{1969Kane}
\begin{align}
 D^b(E)=\frac{k_p}{k_m}\left|\frac{4 k_m k_b }{\varphi\,(k_p^2-k_b^2)^{\frac{1}{2}}(k_m^2-k_b^2)^{\frac{1}{2}}}\right|^2, \label{eq-trans-prob}
\end{align}
where 
\begin{align}
 \varphi= \exp[-\gamma]\exp[ik_b \beta w_b]-\exp[\gamma]\exp[-ik_b \beta w_b], 
\end{align}
and
\begin{align}
 \gamma= \tanh^{-1}(k_b/k_p)+\tanh^{-1}(k_b/k_m). 
\end{align}
As we are assuming plane waves in the plasma and metal regions, $k_p$ and $k_m$ are real. But $k_b$ can be imaginary in the case of tunneling through the sheath-barrier, and it is real in the case of scattering through the barrier. Also, from Eq.(\ref{eq-plasma-e-energy})-Eq.(\ref{eq-energy-conserve}), it is clear that the factors $(k_p^2-k_b^2)^{\frac{1}{2}}$ and $(k_m^2-k_b^2)^{\frac{1}{2}}$ in the denominator of Eq.(\ref{eq-trans-prob}) are independent of $E$ and so they can be taken outside of the integration in Eq.(\ref{eq-biased-J}) when Eq.(\ref{eq-trans-prob}) is substituted in Eq.(\ref{eq-biased-J}). 
\section{I-V Characteristics\label{sec-char}}
In deriving Eq.(\ref{eq-biased-J}), the only assumption made on temperature is that the ionic temperature is close to absolute zero so that the ion current is negligible compared to the electron current. Note that at $V_m=V_f$ i.e., at $V=0$, the current $J$ in Eq.(\ref{eq-biased-J}) is equal to zero. If we assume that the Fermi-Dirac distribution function in Eq.(\ref{eq-biased-J}) approximately has the same shape as in the case of $T_{me}=0$, but without actually setting $T_{me}$ as zero, then the net current becomes
\begin{align}
 J(V)\!=\!\frac{e}{\pi\hbar}\left(\exp\!\left[\frac{eV}{KT_{me}}\right]\!-\!1\right)\!\!\int\limits_{E_{sp}}^{E_{Fm}-eV} \!\! dE\, D^b(E). 
 \label{eqbiasedJapproxo}
\end{align}
At $V_m=V_{sp}$, there is no sheath \cite{conde2011introduction} and so $D^b=1$ which gives the current in the ideal case of Eq.(\ref{eqbiasedJapproxo}) as 
\begin{align}
 J(V_{sp})\!=\!\frac{e}{\pi\hbar}\!\left(\!\exp\!\left[\frac{e(V_{sp}-V_f)}{KT_{me}}\right]\!-\!1\!\right) (E_{Fm}+eV_f). \label{eq-J-at-Vm=Vsp}
\end{align}

For plane waves in the plasma region,  the energy $E$ of a plasma electron should be above $E_{sp}$ in Eq.(\ref{eqbiasedJapproxo}). Also, if the upper limit in the integration in Eq.(\ref{eqbiasedJapproxo}) $E_{Fm}-eV> E_{sp} + \alpha e(\phi_m+\phi_b-V)$, then the electrons with energies greater than $E_{sp}+ \alpha e(\phi_m+\phi_b-V)$ can jump over the sheath-barrier, else for energies less than $E_{sp}+ \alpha e(\phi_m+\phi_b-V)$,  there will be only tunneling through the sheath-barrier. 

\section{Concluding Remarks}
For the bias voltage $V=0$ with respect to the floating potential $V_f$, Eq.(\ref{eq-biased-J}) and Eq.(\ref{eqbiasedJapproxo}) correctly give $J=0$. At this floating potential, $S_b=S_{bf}$. Also, for the probe voltage $V_m=V_{sp}$, since there is no sheath, $S_b=0$, which suggests an expression for $S_b$ as $S_b=S_{bf}-F(y)S_{bf}$, where $y=\left(\frac{V_m-V_f}{V_{sp}-V_f}\right)$ and the function $F(y)$ is such that $F(0)=0$ and $F(1)=1$. 
%For the probe voltage higher than $V_{sp}$, Eq.(\ref{eq-biased-J-approx1}) is different from Eq.(\ref{eq-J-at-Vm=Vsp}). 
A detailed analysis of Eq.(\ref{eq-biased-J}) and Eq.(\ref{eqbiasedJapproxo}) at all bias voltages and the appropriate fixing of the parameters $\alpha$ and $\beta$ will be reported elsewhere. 

%\section*{Data Availability}
%
%No data were generated or used in this work and the work is entirely analytical in nature.
%
%\section*{Authors Declaration}
%
%The authors have no conflicts of interest to disclose. 
%\nocite{*}
%
\section*{References}
\bibliography{PMJunctionQT}% Produces the bibliography via BibTeX.

\end{document}